# Topotactic phase transition in epitaxial La$_{0.7}$Sr$_{0.3}$MnO$_{3-\delta}$ films induced by oxygen getter assisted thermal annealing

*Chenyang Yin[†, ‡, \*], Lei Cao [Θ, ‡], Xue Bai[Δ], Suqin He[±, ‡], Hengbo Zhang[‡], Tomas Duchon[§], Felix Gunkel[±], Yunxia Zhou[#], Mao Wang[#, &], Anton Kaus[±], Janghyun Jo[Δ], Rafal E. Dunin-Borkowski[Δ], Shengqiang Zhou[#], Thomas Brückel[‡], Oleg Petracic[‡, †, \*]*

†: Heinrich Heine University Düsseldorf, Faculty of Mathematics and Natural Sciences, Düsseldorf, 40225, Germany

‡: Jülich Centre for Neutron Science (JCNS-2), JARA-FIT, Forschungszentrum Jülich GmbH, Jülich, 52428, Germany

Θ: School of Advanced Materials, Peking University, Shenzhen Graduate School, Shenzhen, 518055, China

Δ: Ernst Ruska-Centre for Microscopy and Spectroscopy with Electrons (ER-C), JARA-FIT, Forschungszentrum Jülich GmbH, Jülich, 52428, Germany

±: Peter Grünberg Institut (PGI-7), JARA-FIT, Forschungszentrum Jülich GmbH, Jülich, 52428, Germany

§: Peter Grünberg Institut (PGI-6), JARA-FIT, Forschungszentrum Jülich GmbH, Jülich, 52428, Germany

#: Institute of Ion Beam Physics and Materials Research, Helmholtz-Zentrum Dresden-Rossendorf (HZDR), Dresden, 01328, Germany

&: Laboratory of Micro-Nano Optics, College of Physics and Electronic Engineering, Sichuan Normal University, Chengdu, 610101, China

*Corresponding authors:

c.yin@fz-juelich.de

o.petracic@fz-juelich.de

ABSTRACT:

Oxygen vacancies play a crucial role in controlling the physical properties of complex oxides. In $La_{0.7}Sr_{0.3}MnO_{3-\delta}$, the topotactic phase transition from Perovskite (PV) to Brownmillerite (BM) can be triggered e.g. via oxygen removal during thermal annealing. Here we report on a very efficient thermal vacuum annealing method using aluminum as an oxygen getter material. The topotactic phase transition is characterized by X-ray Diffraction which confirms a successful transition from PV to BM in La$_{0.7}$Sr$_{0.3}$MnO$_{3-\delta}$ thin films grown via physical vapor deposition. The efficiency of this method is confirmed using La$_{0.7}$Sr$_{0.3}$MnO$_{3-\delta}$ micron-sized bulk powder. The accompanying transition from the original Ferromagnetic (FM) to an Antiferromagnetic (AF) state and the simultaneous transition from a metallic to an insulating state is characterized using Superconducting Quantum Interference Device (SQUID)-magnetometry and Alternating Current (AC) resistivity measurements, respectively. The near surface manganese oxidation states are probed by synchrotron X-ray Absorption Spectroscopy. Moreover, X-ray Reflectivity, Atomic Force Microscopy and Scanning Transmission Electron Microscopy reveal surface segregation and cation redistribution during the oxygen getter assisted annealing process.

INTRODUCTION:

Complex oxides have been the subject of extensive research in recent decades, owing to their wide-ranging potential applications [1-5]. A practical strategy to control the properties of complex oxides is by adjusting the concentration of oxygen vacancies. The movement and diffusion of oxygen vacancies can provide a resistive switching mechanism or introduce sites for surface reactions, thereby supporting applications such as gas sensors and catalysts [6-17]. In solid oxide fuel cells, oxygen vacancies are crucial for the devices' operation and performance, facilitating oxygen transport in electrolytes and electrodes [18-21].

The extraordinary role of controlling the concentration of oxygen vacancies also extends to spintronic properties of materials. In a multivalent system involving transition metal ions, the oxidation state of the 3d cations is directly related to the oxygen off-stoichiometry. Consequently, controlling the concentration of oxygen directly influences several physical properties such as electronic transport and magnetic properties [22-26].

An oxygen vacancy induced topotactic phase transition from Perovskite (PV, $ABO_3$) phase to the oxygen-vacancy-layered Brownmillerite (BM, $ABO_{2.5}$) phase has been reported in $La_{1-x}Sr_xMnO_{3-\delta}$ [27-31] and in other analogous perovskite systems e.g. $La_{1-x}Sr_xCoO_{3-\delta}$ [32-35], $SrFe_{1-x}Co_xO_{3-\delta}$ [36], $SrFeO_{3-\delta}$ [37] and $SrCoO_{3-\delta}$ [38-40]. Via the (primarily) structural topotactic phase transition, simultaneously a phase transition in the magnetic and electron transport properties can be triggered [32]. The PV-BM topotactic phase transition can hereby be realized using various methods, e.g. thermal annealing [36], thermal vacuum annealing [28-29,32-33], voltage pulse control [27], Ionic Liquid Gating (ILG) [38,41] or Ionic Gel Gating (IGG) [34] and reduction gas annealing [30-31].

In this study, we employ bulk aluminum as an oxygen getter material to increase the efficiency of the thermal vacuum annealing process. The PV ($La_{0.7}Sr_{0.3}MnO_3$) to BM ($La_{0.7}Sr_{0.3}MnO_{2.5}$) transition is successfully achieved and characterized using X-ray Diffraction (XRD). In addition to the thin film sample, a successful phase transition is also observed in micron-sized powder. The physical properties changes are characterized using SQUID-magnetometry and AC resistivity measurements. X-ray Absorption Spectroscopy (XAS) probes the near surface oxidation states of the multivalent manganese ions. The variations of the surface and the interface before and after annealing are investigated using X-ray Reflectivity (XRR) and Atomic Force Microscopy (AFM). The cation stoichiometry is probed via Rutherford Backscattering Spectroscopy (RBS).

These results suggest a controlled PV to BM transition, accompanied by a complex near surface structural rearrangement and cation redistribution. This is confirmed via Scanning Transmission Electron Microscopy (STEM) measurements. Hence, this finding can be of vital importance for the understanding and the function of devices such as those used in catalysis and fuel cells, where the surface and the interface play a crucial role.

RESULTS AND DISCUSSION:

Compared with the Fe or Co systems, it requires relatively high temperatures and prolonged durations to trigger the PV to BM topotactic phase transition using thermal annealing in $La_{0.7}Sr_{0.3}MnO_{3-\delta}$ (LSMO) [28-29]. Here, we employ bulk aluminum as an oxygen getter to increase the efficiency of the thermal annealing process. Epitaxial LSMO films were grown using High Oxygen Pressure Sputtering Device (HOPSD) on $SrTiO_3$ (STO) substrates with (100) orientation. The growth temperature was 800 °C at an oxygen partial pressure of 2 mbar. The detailed growth procedure is illustrated in Experimental Section 1. Selected films, e.g., for the STEM studies, were grown by Pulsed Laser Deposition (PLD) for an improved surface morphology and reduced surface roughness, see Experimental Section 3. The as-prepared films were pre-characterized to ensure a good thin film quality for the subsequent oxygen getter assisted thermal vacuum annealing. Aluminum, being a reactive metal, forms a surface oxide film upon exposure to oxygen. This oxide layer passivates the surface, thereby inhibiting further oxidation. An approximately 3 nm thick surface oxide layer is formed according to the chemical reaction: $4Al+3O_2 \rightarrow 2Al_2O_3$ on the time scale of ∼one day [42].

In addition, the thickness of this oxide layer can be increased to ca. 200 nm at temperatures exceeding 400 °C [43]. In this study we employed regular aluminum foil as getter material. By removing the surface oxide layer by mechanical polishing, the aluminum foil with the exposed bare aluminum surface can act as an efficient oxygen getter material during thermal vacuum annealing. In our study, the as-prepared LSMO films are sealed together with freshly polished aluminum foil in a quartz tube, which is subsequently evacuated to high vacuum of $10^{-5}$-$10^{-6}$ mbar. The sealed tubes are then annealed in a temperature-controlled furnace at 350 °C or 400 °C for 12 hours, respectively. Experimental Section 2 describes the details of the oxygen getter assisted thermal annealing experiment.

XRD is employed for crystal structure characterization before and after annealing. The results are illustrated in Figures 1b and 1c. The as-prepared LSMO film grown via HOPSD, labelled as "PV-LSMO" state, shows (002) Bragg reflections at the scattering vector $Q_Z = 3.261 \pm 0.002$ Å$^{-1}$ which corresponds to an out-of-plane lattice parameter of $3.853 \pm 0.002$ Å. The clear Laue oscillations in the vicinity of the (002) reflection indicate the good crystal quality of the as-prepared thin film. The Reciprocal Space Mapping (RSM) results are shown in Figure 1f. The film and the substrate exhibit the same in-plane lattice parameter, confirming a fully strained film growth.

Subsequently, the films were treated using Al-assisted thermal annealing at various temperatures to trigger the topotactic phase transition. One as-prepared film was annealed at 350 °C for 12 hours. Its (002) reflection shifts to a lower $Q_Z$ position without the emergence of new reflections. This indicates a lattice expansion within the existing PV framework. This expanded PV state is labelled as "E-PV350". At this state the out-of-plane lattice parameter is $3.934 \pm 0.002$ Å. Thus, it shows an expansion by 2.1 % compared with the as-prepared state, i.e. $3.853 \pm 0.002$ Å. This expansion can be interpreted as a result of the oxygen vacancies $V_O^{\ddot{}}$ created. The presence of vacancies results in a lower oxidation state of the Mn ions in order to compensate for the charge imbalance due to the missing O$^{2-}$ ions. These cations have a larger ionic radius. In addition, the lattice parameter increases due to the electrostatic repulsion of (in absence of some oxygen ions) unscreened cations [44-45].

To improve transition efficiency, the temperature was increased to 400 °C for another as-prepared film, while keeping the annealing time at 12 hours. The corresponding XRD pattern shows clear superlattice Bragg peaks, which indicate an oxygen-vacancy-layered structure and are considered as characteristic fingerprints of the BM structure (Figures 1a and 1c) [46]. The label "BM400" is assigned to this state in the subsequent discussion. Supporting information S1 shows the detailed sample information.

The surface and in-depth information is probed via XRR (Figure 1d and 1e) and AFM. We fit the XRR pattern using a three-layer model using the software GenX [47]. Thickness and roughness for each layer and the corresponding Scattering Length Density (SLD) are determined from fitting (Figure 1e). The supporting information S2 displays the fitting models and results. We find that the central layer SLD of E-PV350 is decreased from $4.64 \pm 0.05 \times 10^{-5}$ Å$^{-2}$ to $4.28 \pm 0.05 \times 10^{-5}$ Å$^{-2}$ ($-7.8 \pm 1.5$ %). The reduction in SLD can be attributed to the oxygen vacancy induced lattice expansion. We also observe a decrease in film thickness, i.e. $41.9 \pm 0.5$ nm to $37.5 \pm 0.5$ nm ($-10.5 \pm 1.7$ %). This may be a counter intuitive result and can be attributed to, e.g., the breakdown of the crystal structure and the redistribution of cations, which will be discussed in more detail below.

The three-layer-model exhibits a poor fitting quality when applied to the BM400 sample. Hence, we assume the formation of a new layer near the surface. Since X-rays exhibit a low sensitivity to light elements, e.g., oxygen, this additional layer is likely due to a cation redistribution. The complex situation inside the BM400 still cannot be described by a simple four-layer model, resulting in imperfect fitting results as well. In this four-layer-model the main LSMO layer near the substrate shows a SLD of $3.64 \pm 0.05 \times 10^{-5}$ Å$^{-2}$ ($-21.6 \pm 1.5$ %). This SLD decrease, i.e., $-21.6 \pm 1.5$ %, is much larger than the oxygen vacancy induced SLD change from bulk La$_{0.7}$Sr$_{0.3}$MnO$_3$ to bulk La$_{0.7}$Sr$_{0.3}$MnO$_{2.5}$ ($-11.3$ %). This implies a strong cation-related redistribution and segregation. The bulk La$_{0.7}$Sr$_{0.3}$MnO$_3$ and La$_{0.7}$Sr$_{0.3}$MnO$_{2.5}$ SLD values were calculated using the data from ICSD-50717 (ICSD release 2024.1) and from ICSD-166141 (ICSD release 2024.1), respectively [46, 48]. Supporting information S3 illustrates the results of the samples annealed at 300 °C with different aluminum content and at 450 °C. Supporting information S4 shows the stability of the E-PV350 and BM400.

To further test the Al-assisted annealing method, micron-sized LSMO bulk powder was used. Compared with the LSMO thin films, the significantly larger diffusion paths in the powder particles should significantly inhibit the topotactic phase transition. However, a PV to BM topotactic phase transition is successfully achieved using Al-assisted vacuum annealing at 550 °C for 20 hours. For details, please see Supporting information S7. This annealing condition is still lower or comparable to the conventional vacuum annealing conditions applied to thin films [28-29]. This again confirms the efficiency of the Al-assisted thermal annealing.

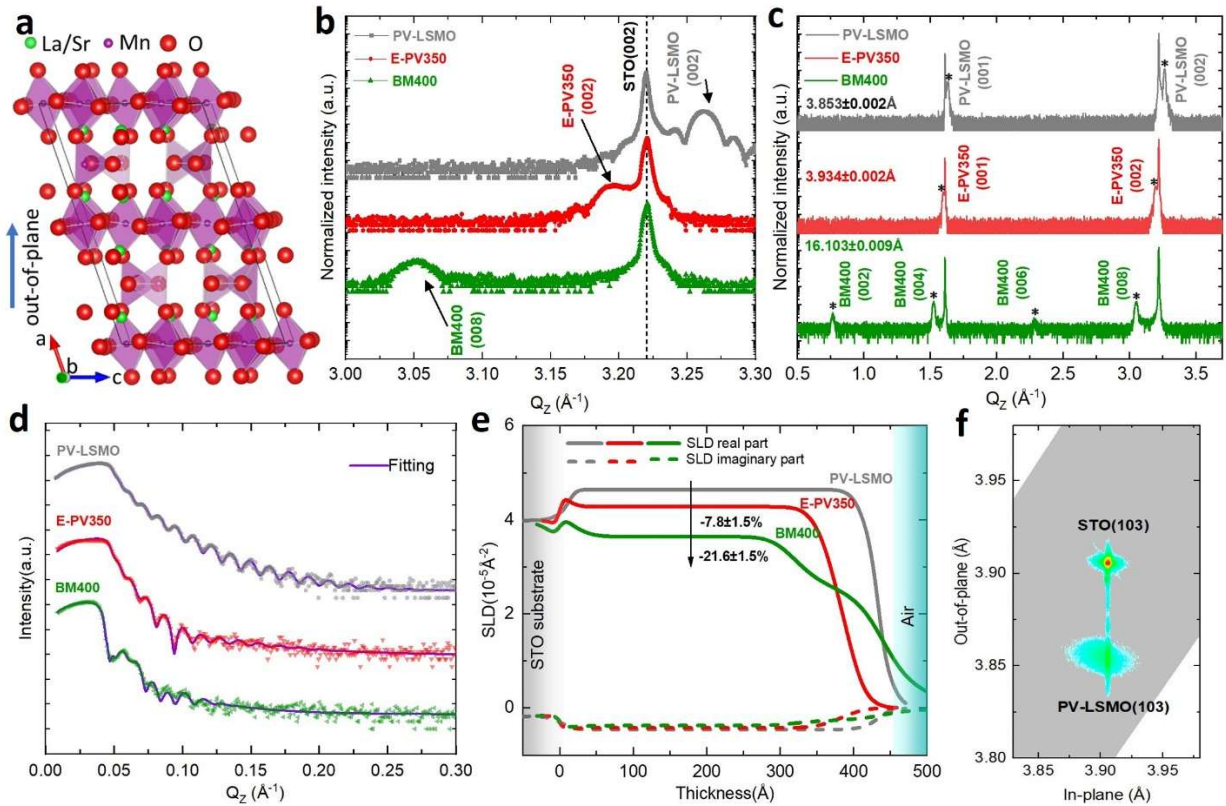

Figure 1. (a) Crystal structure schematics of the BM phase, plotted using the software VESTA [49], data taken from ICSD-166141 (ICSD release 2024.1) [46]. (b) XRD patterns of the as-prepared (grey) and the annealed LSMO films, annealed for 12 hours at 350 °C (red) and for 12 hours at 400 °C (green), respectively. XRD measured at room temperature with the $Q_Z$ range between 3 and 3.3 Å$^{-1}$. The black arrows indicate the thin film (002) Bragg reflection. (c) XRD patterns with $Q_Z$ range between 0.5 and 3.75 Å$^{-1}$. Stars indicate the film Bragg reflections. The out-of-plane lattice parameters as calculated are shown. (d) XRR patterns of the as-prepared (grey) and annealed samples (red and green) recorded at room temperature. The fitting curves using the software GenX are shown in purple. (e) Real (Re.) and imaginary (Im.) SLD profiles as derived from the fitting. The thickness on the horizontal axis refers to the distance to the substrate, i.e. 0 refers to the position on STO substrate surface. (f) RSM pattern of the as-prepared LSMO film. The (103) Bragg reflection is probed for the thin film and for the STO substrate to compare the in-plane strain state. A fully strained film on the substrate is hence confirmed.

AFM provides information about the thin film surface morphology (Figure 2 a-f). The PV-LSMO film shows a surface with granular morphology with an average height of 5 nm. The Root Mean Square (RMS) roughness is calculated to be 1.9 ± 0.2 nm. After annealing, the E-PV350 exhibits a comparable RMS of 1.4 ± 0.2 nm and a less granular surface structure. For the sample BM400 which is obtained with annealing at higher temperatures i.e. at 400 °C, the RMS increased to 2.7 ± 0.2 nm together with the formation of larger lateral structures, which hints at a surface segregation.

The oxygen vacancies as well as the induced structural changes have an impact on the magnetic exchange interactions between manganese ions and hence on the overall magnetic behavior and the electric transport properties. To elucidate the role of oxygen vacancies on the magnetic and transport properties, SQUID-magnetometry and resistivity measurement were performed. For the PV-LSMO, the average manganese oxidation state is calculated to be $Mn^{3.3+}$ by the stoichiometry of $La_{0.7}Sr_{0.3}MnO_{3-\delta}$ with the assumption of oxidation states for $La^{3+}$, $Sr^{2+}$ and $O^{2-}$ and $\delta$ close to zero after the growth under high oxygen pressure [28]. The magnetic exchange interactions via oxygen bridges among manganese ions with different oxidation states, e.g. $Mn^{3+}$ and $Mn^{4+}$, promotes double-exchange interactions and thus overall ferromagnetic behavior [50]. As illustrated in Figure 2g, the ferromagnetic behavior is confirmed from the magnetization vs. temperature Field Cooling (FC) curve. The Curie temperature is determined to be $327 \pm 2$ K. The kink at $105 \pm 2$ K is due to the influence of the STO structural phase transition from cubic to tetragonal onto the film magnetic anisotropy [51]. In the hysteresis loop (Figure 2i), the PV-LSMO sample displays a rectangular shaped loop indicating that the magnetization reversal occurs by domain wall motion. The saturation magnetization is determined to be $3.24 \pm 0.05$ $\mu_B$/Mn with a coercive field of $14 \pm 1$ mT.

Via Al-assisted thermal annealing, oxygen vacancies are introduced into the perovskite structure. The electron hopping between Mn sites is interrupted by the missing oxygen and thus suppresses double-exchange interactions. This gradual breakdown of the double-exchange paths can be considered as analogous to the percolation model at first glance. However, the variation in the charge environment surrounding the manganese ions due to the oxygen vacancies leads to a lower manganese oxidation state. The reduction of the manganese oxidation state favors super-exchange interactions between manganese ions which in turn lead to antiferromagnetic coupling. Consequently, a more complex percolation model including both the interruption of exchange paths and the local change of the oxidation state of Mn ions needs to be considered. Such a numerical study is, however, beyond the scope of this manuscript.

For E-PV350, the emergence of a peak at 46 ± 2 K in the Zero-Field Cooling (ZFC) curve indeed suggests antiferromagnetic behavior (Figure 2h). In addition, the *M* vs *H* curve of this sample shows a very small saturation magnetization and a small remanent magnetization which also hints toward antiferromagnetic behavior. Interestingly, the BM400 sample exhibits straight-line shaped ZFC and FC curves without any significant magnetic signatures. This might hint toward a different magnetic behavior.

Measurements of electric transport properties are shown in Figure 2j. The PV-LSMO sample shows an insulator to metal transition near the Curie temperature as expected [52]. Upon annealing, the double-exchange paths are suppressed, and super-exchange interactions become preferred which leads to an overall insulating behavior. A transition into insulating state within the PV phase is also consistent with previous reports on manganites and cobaltites [28, 32].

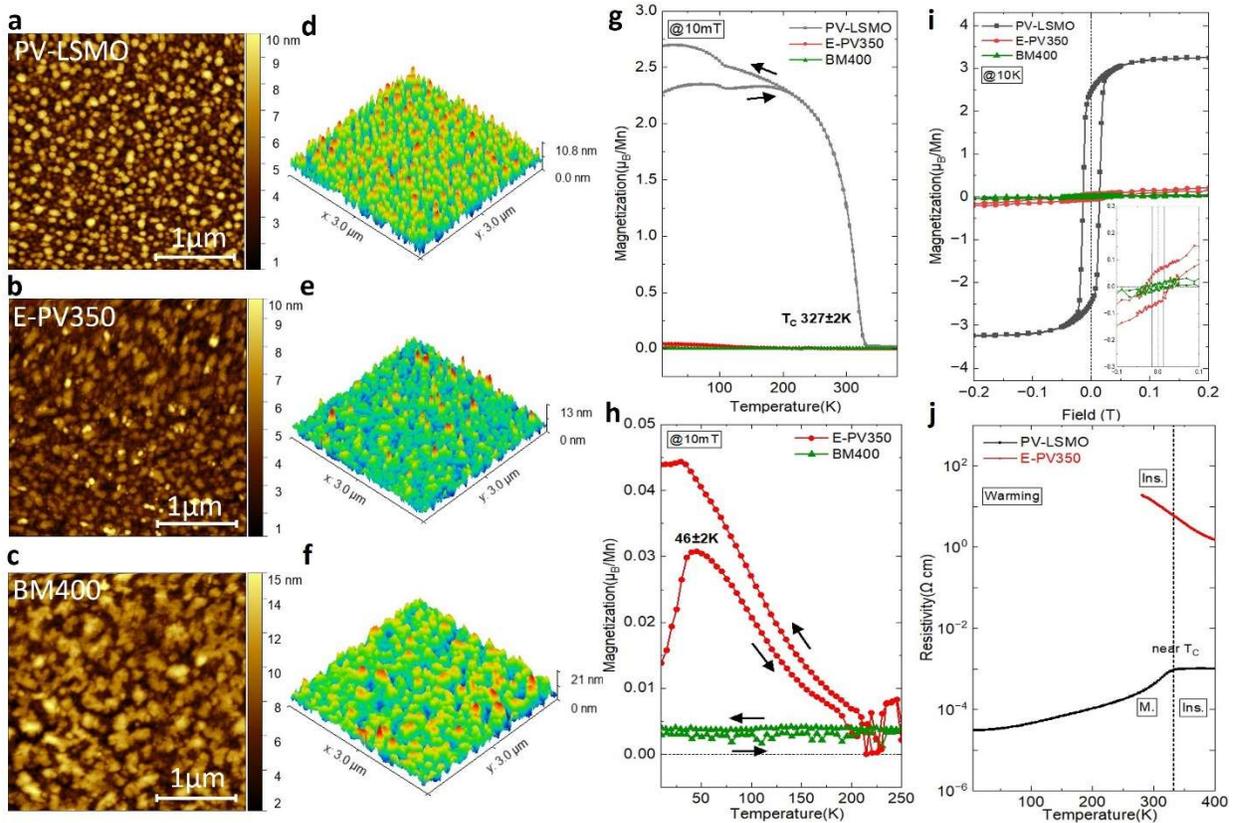

Figure 2. (a-c) AFM images: 2D morphology of the as-prepared and the annealed films. Measurements were performed at room temperature. (d-f) Corresponding 3D morphology. (g) Magnetization vs. temperature measurements of the as-prepared and the annealed films. ZFC and FC curves are displayed. Measurements were conducted at a magnetic field of 10 mT. The arrows denote the temperature scan direction during measurement. (h) Zoom-in of the ZFC and FC curves of the annealed films for a better comparison. (i) Magnetization vs. field hysteresis loops measured at 10 K. (j) Resistivity vs. temperature measurements. The curves were measured without magnetic field during the warming process. The dotted line represents the transition from conductor to insulator for the as-prepared film.

In order to explain the above-described magnetic behavior, the detailed electronic structure of the magnetic ions is of critical importance. Hence, high-energy-resolution XAS was employed. In XAS measurements, two commonly used detection methods are the Total-Electron-Yield (TEY) signal and the Fluorescence-Yield (FY) signal. Here, the TEY signal is shown due to its high surface sensitivity, which can probe a depth range of a few nanometers due to exponential intensity decay. The oxidation state of manganese can be deduced from the peak position and the line shape of the manganese L-edge (Figure 3a). By comparing the XAS data with data from reference samples [53-54], $Mn^{3+}$ is shown to dominate in the PV-LSMO sample with some $Mn^{4+}$ contributions corresponding to the average valence of $Mn^{3.3+}$ as expected for a fully oxidized as-grown sample. This result corroborates the ferromagnetic behavior induced by the double-exchange bridges between $Mn^{3+}$ and $Mn^{4+}$. For the E-PV350, the shape of $L_2$ edge becomes more oriented to that of $Mn^{2+}$ implying an increase of the $Mn^{2+}$ contribution (see reference $Mn^{2+}$ curve) thus promoting super-exchange interactions among $Mn^{2+}$ leading to antiferromagnetic behavior.

The larger $L_3$ to $L_2$ branching ratio in the BM400 sample indicates an increased amount of high-spin states and thus a larger amount of $Mn^{2+}$ species [54-55]. In this sample only the features of $Mn^{2+}$ are observed near the sample surface. In addition, the oxygen K-edge is shown in Figure 3b. The K-edge originates from the electron excitation from the oxygen 1s orbitals to the conduction band i.e. to the unoccupied hybridization orbitals between oxygen 2p orbitals and cation orbitals. The hybridization regions for different cation orbitals are labelled in Figure 3b [29, 56]. This oxygen K-edge can indirectly probe the occupancy of manganese 3d orbitals. After annealing, the intensity of the $e_g\uparrow$ peak decreases and of the $t_{2g}\downarrow$ peak increases. This indicates fewer unoccupied manganese 3d $e_g\uparrow$ hybridized orbitals available, which corresponds to more electrons in the manganese 3d orbitals compared with the as-prepared state, thus reflecting a lower oxidation state of manganese. This result corresponds well to the results from the manganese L-edge. In addition, the loss of structural features at larger photon energies (the O 2p, La 5d, Sr 4d, Mn 4sp range) may further indicate a loss of crystallinity in the near-surface region [57]. The lanthanum M-edge is displayed in Figure 3c. The oxidation state of lanthanum stays unchanged after annealing according to the line shape. However, TEY-XAS probes regions very near to the surface. The depth-resolved oxidation state mapping of a similar sample is shown in Figure 4.

In addition to oxygen vacancy formation, cations in perovskites can also diffuse and segregate during annealing and form e.g. Ruddlesden-Popper phases [58-59]. In order to characterize the thin film cation stoichiometry and the corresponding depth-distribution, Rutherford Backscattering Spectroscopy was employed. This method is not sensitive to light elements, such as oxygen. Hence, only the stoichiometry of La, Sr and Mn can be determined using this method. The La + Sr is normalized to 1 during the stoichiometry determination. The fitting is performed using a single LSMO layer model via the software SIMNRA [60-61]. The results show a clear overall manganese deficiency (-19.6%) in the E-PV350 sample compared with the PV-LSMO. In the BM400, this manganese deficiency is enhanced and reaches -43.3%. In addition, the peak shape of manganese in the BM400 indicates a different manganese concentration near the surface which matches the findings from XRR and AFM measurements suggesting a surface segregation and cation redistribution.

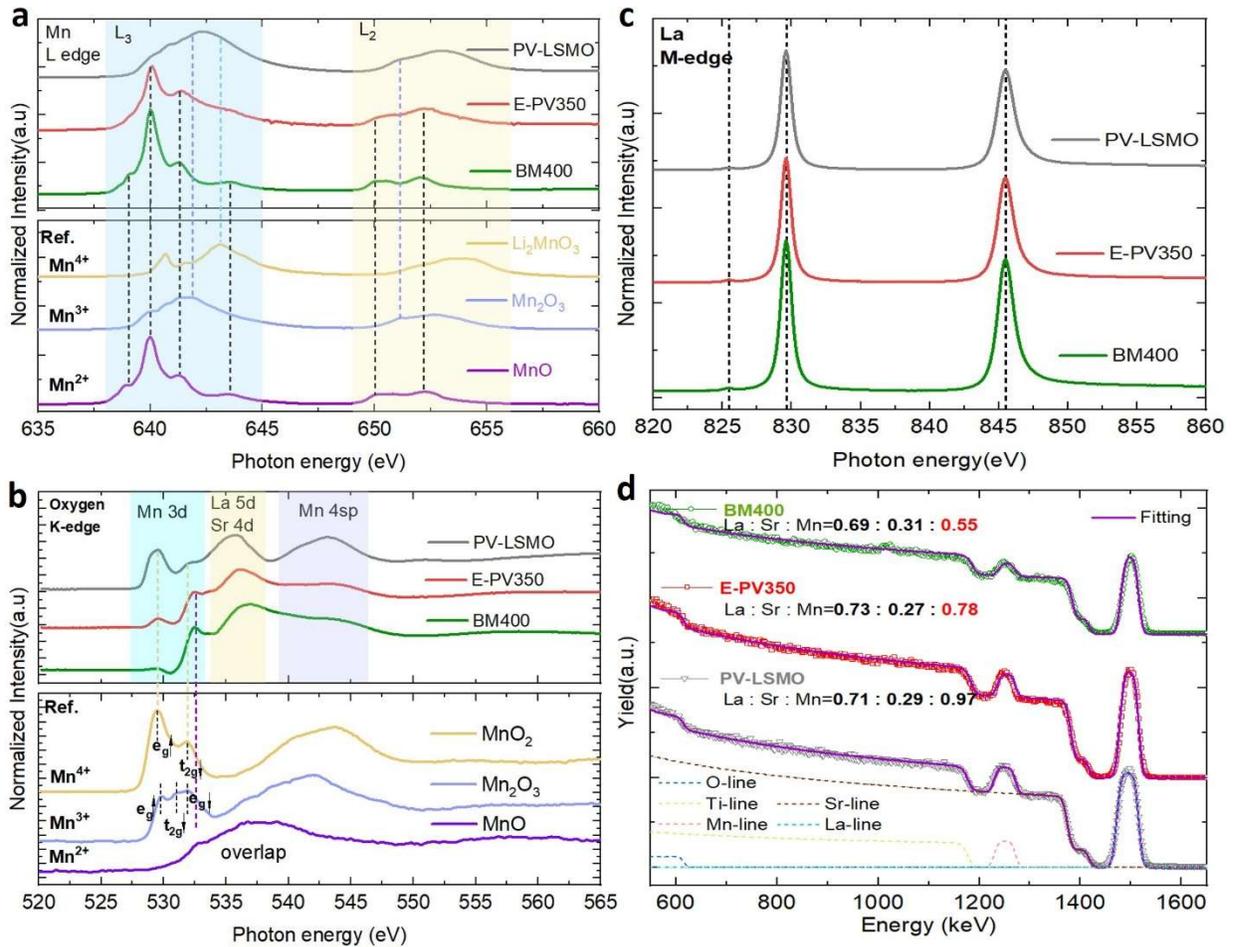

Figure 3. Oxidation state and stoichiometry characterization, measured at room temperature using XAS and RBS, respectively. (a) XAS manganese $L_{2,3}$-edges spectra of the as-prepared and the annealed LSMO films on STO substrate, annealed for 12 hours at 350 °C and at 400 °C, respectively. The reference manganese oxidation data is reprinted with permission from [53-54]. Peak features are labeled with dashed lines. (b) Corresponding XAS oxygen K-edge spectra. The different manganese $e_g$ and $t_{2g}$ orbital feature peaks are labeled. (c) XAS lanthanum M-edge spectra. (d) RBS spectra of the as-prepared and the annealed LSMO films. The dashed lines represent the contribution of each element in the as-prepared film. The calculated cation stoichiometry determined by fitting the measurement curves is displayed.

The results of XRR, AFM, XAS, and RBS suggest a near surface structural transformation and accompanied by cation redistribution. To investigate this phenomenon in detail, an LSMO thin film was prepared using PLD, which exhibits an improved surface quality, as illustrated in Supporting information S8. This minimizes the influence of surface roughness, enabling a more accurate study of near surface phenomena. The annealing experiment of BM400 is repeated on this sample. Clear BM superlattice peaks emerge and are shown in Supporting information S8. This sample is labeled "PLD400" for further reference. STEM is subsequently employed for a detailed characterization of the film structure, as demonstrated in Figure 4.

Figure 4a shows Energy Dispersive X-ray Spectroscopy (EDS) elemental mapping. A clear manganese segregation is observed near the thin film surface. Below the segregation region, a two-layer structure is evidenced. The top layer, near the surface, exhibits an amorphous structure and shows significantly low concentrations of lanthanum, manganese, and oxygen. The depth-dependent elemental profile is displayed in Figure 4b. Manganese is found to be more deficient compared to other cation elements, i.e., Sr and La. This is consistent with the RBS data shown in Figure 3d. The high resolution HAADF image shows a clear atomic ordering for the second layer (Figure 4c) with the same orientation as the substrate. From the integrated height profile (Figure 4d), the alternating layers with high and low concentration of oxygen vacancies are observed as expected for a BM structure. This layer is considered to be the major contribution to the BM-LSMO signal as found from XRD measurements. The EELS oxidation state mapping is shown in Figure 4e and 4f. From the integrated EELS signal, a shift of the manganese oxidation state is found depending on the depth. From the manganese $L_3$-edge peak position, we conclude that the low oxidation state of Mn, i.e., close to $Mn^{2+}$, dominates in the manganese segregated surface region and also in the amorphous layer [62]. However, in the crystalline second layer, the oxidation state is close to $Mn^{3+}$. The observed AF magnetic behavior for sample E-PV350 and the unusual magnetic behavior in BM400 might hint at a novel type of spin structure involving mainly $Mn^{3+}$ species and very likely only super-exchange interactions.

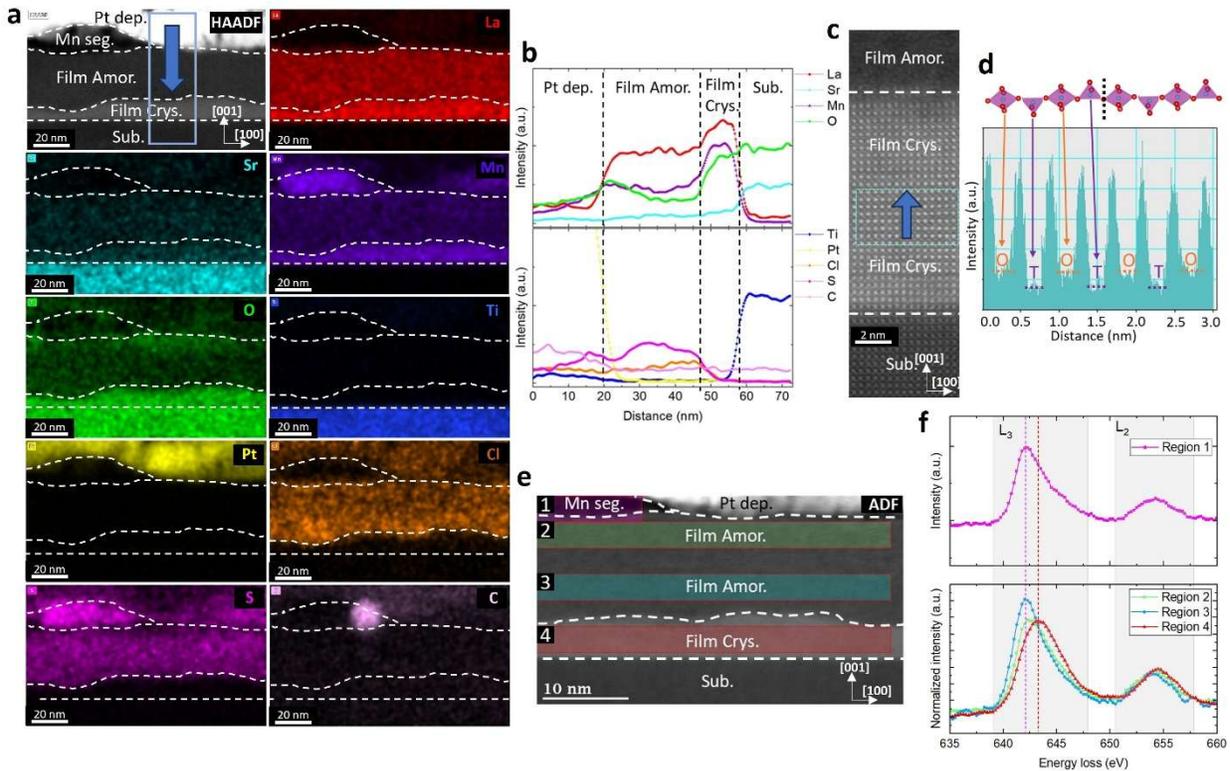

Figure 4. STEM measurement on the 400 °C 12 hours annealed PLD sample. Data analyzed using the Gatan DigitalMicrograph software. (a) HAADF images and EDS elemental mapping. The mapping indicates clearly regions with varying elemental distributions within the annealed film. The white dotted lines are used to mark the boundaries between regions. The net intensity data is shown here. For the origin of Pt, Cl, S, and C please see Supporting information S8. (b) Elemental

intensity profile along the blue arrow depicted in (a). The data shown is the integrated signal within the blue box. (c) Zoom-in of the high resolution HAADF image. The upper dark region is the amorphous layer. The middle region belongs to the crystallized BM-LSMO. The bottom region is the STO substrate. (d) Profile along the blue arrow in (c). A clear oxygen octahedra and tetrahedra alternating structure is observed. (e) ADF image in the EELS mapping. Four regions are selected. Region 1: Manganese segregated region. Region 2: Amorphous region near surface. Region 3: Amorphous region in the middle of the film. Region 4: Crystallized BM-LSMO region. (f) Integrated EELS signal of the different regions in (e). The signals from regions 2, 3, and 4 are normalized. For Region 1, likely due to a significant thickness difference along the [010] direction, normalization is not possible and only the deconvoluted signal is shown.

CONCLUSION:

We demonstrate that the PV to BM topotactic phase transition in LSMO thin films can be successfully triggered by thermal vacuum annealing using aluminum as an oxygen getter at comparably lower annealing temperatures of 400 °C. The efficiency of this method is confirmed for both thin film samples and micron-sized powder. We confirm that the LSMO films undergo a transition for the magnetic and electric transport properties from a FM to an AF state and simultaneously from a metallic to an insulating state as the manganese oxidation state decreases. The results from XRR, AFM, and RBS suggest structural transformations near the film surface accompanied by cation redistribution. STEM results reveal a clear manganese segregation at the surface and subsurface cation redistribution, where an amorphous top layer is formed. The near substrate layer shows a BM-LSMO crystal structure. This method can be employed to study oxygen vacancy induced topotactic phase transitions also in similar oxide systems especially where extremely high temperatures would be required.

ASSOCIATED CONTENT:

EXPERIMENTAL SECTION:

1. HOPSD: Epitaxial $La_{0.7}Sr_{0.3}MnO_{3-\delta}$ thin films were initially grown using HOPSD [63]. A stochiometric $La_{0.7}Sr_{0.3}MnO_3$ sputtering target from the company EVOCHEM was used for the HOPS deposition. The $SrTiO_3$ substrates with (100) orientation from the company SHINKOSHA were pre-annealed at 950 °C under an oxygen partial pressure of 2 mbar for 30 minutes to remove surface adsorbents [64]. Simultaneously a target pre-sputtering at a plasma power of 100 W was performed for 160 minutes to remove adatoms on the target. Subsequently, the plasma power was increased to 120 W while the substrate temperature was decreased to 800 °C for thin film growth. Thin films with a thickness of 40 nm were achieved with a growth rate of 40 nm/hour. The prepared samples were then cooled at 2 mbar from the growth temperature to room temperature with a rate of 5 °C/min.

2. Oxygen getter assisted thermal annealing: Commercial lab Aluminium foil was polished using sandpaper P400 to remove the surface oxide layer. Subsequently the polished Aluminium foil with mass of 0.5 g was sealed together with the LSMO film (with a lateral size of ca. 5 mm×5 mm) in a quartz tube at high vacuum. Quartz wool was placed between the Aluminium foil and the LSMO film to avoid direct contact. The tube volume is ca. 100 $cm^3$. The annealing experiments of the sealed tubes were conducted in a temperature-controlled furnace from the company Nabertherm.

3. PLD: The subsequent high-quality epitaxial $La_{0.7}Sr_{0.3}MnO_{3-\delta}$ thin film with a thickness of 30 nm was produced using a PLD system from the company Twente Solid State Technology B.V. A KrF excimer laser (LPX 300) from the company Lambda Physik was used with a wavelength of 248 nm. The SrTiO$_3$ (100) substrate from the company SHINKOSHA was pre-annealed at 850 °C in the PLD chamber under 0.24 mbar oxygen. The growth followed at the same temperature and pressure. After growth, the sample was cooled down at the growth pressure at a rate of 10 °C/min.

4. XRD and XRR: XRD and XRR measurements were performed using the Bruker D8 Advance instruments at JCNS-2 and at PGI-7. Cu K$_{\alpha 1}$ radiation with an X-ray wavelength of 1.54 Å was used for characterization. The D8 at JCNS-2 is equipped with two Göbel mirrors to achieve a highly collimated beam for precise XRR scans at lower angles. In addition, a channel-cut monochromator is used for accurate XRD scans at higher angles. The D8 at PGI-7 is equipped with a single Göbel mirror for both XRR and XRD scans.

5. AFM: The topographic information of thin film samples was collected by an Agilent 5400 atomic force microscope under the alternating current (AC) tapping mode. The type HQ:NSC15/Al BS AFM silicon probe coated with Aluminum as the reflection layer is from the MikroMasch company.

6. SQUID-magnetometry: Macroscopic magnetic properties i.e. ZFC curves, FC curves and hysteresis loops were measured via a Superconducting Quantum Interference Device (SQUID) magnetometer MPMS XL from the company Quantum Design.

7. PPMS: Thin film resistivity vs. temperature behaviours were characterized by a Physical Properties Measurement System (PPMS) from Quantum Design via the van der Pauw method using 4 contacts.

8. RBS: RBS was measured at HZDR in Dresden. 4-He$^+$ ions are accelerated by the Rossendorf van de Graaff accelerator to 1.7 MeV. The detector is placed at the backscattering angle of 170°.

9. XAS: High resolution XAS was measured at the synchrotron BESSY in Berlin at the beamline UE56/1-SGM. The energy probing range covers Mn $L_{2,3}$-edge, O K-edge and La M-edge. TEY signals were collected. The data normalization follows: (1) Normalization by the background current. (2) The background shift from 625 eV to 670 eV for manganese $L_{2,3}$-edges and from 520 eV to 570 eV for oxygen K-edge is normalized to 1. (3) The manganese $L_{2,3}$-edges data is laterally shifted by +1.8 eV to match the reference data. The oxygen K-edge data is laterally shifted by +0.237 eV to match the reference data.

10. STEM: The FEI Titan G280-200 ChemiSTEM at ERC was employed for HAADF, ADF, EDS, and EELS data collection. The high brightness Schottky type field emission electron gun (FEI X-FEG) operated at 200 kV. The Fischione Model 300 angular dark-field detector recorded the HAADF and ADF images. EDS and EELS were measured using an in-column Super-X EDX spectroscopy unit (ChemiSTEM technology) and a Gatan Enfinium ER 977 post-column energy filter system. For EELS, the energy range 610-630 eV is chosen as background. Plural scattering is removed. The data for regions 2, 3, and 4 are normalized using the integral between 635 eV and 660 eV.

AUTHOR INFORMATION

**Corresponding Author**

* Oleg Petracic

   o.petracic@fz-juelich.de

* Chenyang Yin

   c.yin@fz-juelich.de

**Present Addresses**

**Author Contributions**

**Funding Sources**

ACKNOWLEDGMENT

Chenyang Yin gratefully thanks Dr. Ulrich Rücker, Dr. Emmanuel Kentzinger, Jianwei Ye, Dr. Connie Bednarski-Meinke, Dr. Shibabrata Nandi, Dr. Moritz Weber, Dr. Mai Hussein for help and discussions. Benjamin Reineke is acknowledged for providing SEM at JCNS-1. Chenyang Yin acknowledges Jörg Perßon, Andreas Schwaitzer, Berthold Schmitz, Frank Gossen for technical support. Lei Cao and Yunxia Zhou were partially supported by German Research Foundation (Grant No. ZH 225/10-1). ErUM-Pro is acknowledged for supporting large-scale facilities.

ABBREVIATIONS

ADF, Annular Dark-Field; AF, Antiferromagnetic; AFM, Atomic Force Microscopy; BM, Brownmillerite; EDS, Energy Dispersive X-ray Spectroscopy, EELS, Electron Energy Loss Spectroscopy; FC, Field Cooling; FM, Ferromagnetic; FY, Fluorescence-Yield; HAADF, High-Angle Annular Dark-Field; HOPS, High Oxygen Pressure Sputtering; ILG, Ionic Liquid Gating; IGG, Ionic Gel Gating; LSMO, $La_{0.7}Sr_{0.3}MnO_{3-\delta}$; PPMS, Physical Properties Measurement System; PV, Perovskite; RBS, Rutherford Backscattering Spectroscopy; RMS, Root Mean Square; RSM, Reciprocal Space Mapping; SLD, Scattering Length Density; SQUID, Superconducting Quantum Interference Device; STEM, Scanning Transmission Electron Microscopy. STO, $SrTiO_3$; TEY, Total-Electron-Yield; XAS, X-ray Absorption Spectroscopy; XRD, X-ray Diffraction; XRR, X-ray Reflectivity; ZFC, Zero-Field Cooling.